\documentclass[twocolumn]{revtex4}%
\usepackage{amsfonts}
\usepackage{amsmath}
\usepackage{amssymb}
\usepackage{graphicx}%
\begin{document}
\title{Searching for the Slater Transition in the Pyrochlore Cd$_{2}$Os$_{2}$O$_{7}$
with Infrared Spectroscopy}
\author{W.J. Padilla}
\affiliation{Department of Physics University of California San Diego, La Jolla, CA 92093-0319}
\author{D.N. Basov}
\affiliation{Department of Physics University of California San Diego, La Jolla, CA 92093-0319}
\author{D. Mandrus}
\affiliation{Solid State Division, Oak Ridge National Laboratory, Oak Ridge, TN 37831}

\begin{abstract}
Infrared reflectance measurements were made on the single crystal pyrochlore
Cd$_{2}$Os$_{2}$O$_{7}$ in order to examine the transformations of the
electronic structure and crystal lattice across the boundary of the metal
insulator transition at $T_{MIT}=226K$. All predicted IR active phonons are
observed in the conductivity over all temperatures and the oscillator strength
is found to be temperature independent. These results indicate that charge
ordering plays only a minor role in the MIT and that the transition is
strictly electronic in nature. The conductivity shows the clear opening of a
gap with $2\Delta=5.2k_{B}T_{MIT}$. The gap opens continuously, with a
temperature dependence similar to that of BCS superconductors, and the gap
edge having a distinct $\sigma(\omega)\thicksim\omega^{1/2}$ dependence. All
of these observables support the suggestion of a Slater transition in Cd$_{2}%
$Os$_{2}$O$_{7}$.

\end{abstract}
\maketitle

Mott's classic paper\cite{Mott49} published over half a century ago has
triggered extensive research on correlated electron systems which undergo a
metal insulator transition (MIT). Both Mott and Hubbard\cite{Hubbard63} have
suggested that for systems at half filling, the Coulomb repulsion between
electrons could split the band, thus producing an insulator. Alternatively,
Slater in 1951 suggested that antiferromagnetic order alone could produce an
insulator by a doubling of the magnetic unit cell\cite{Slater51}. While there
are numerous examples of Slater / spin-density-wave (SDW) insulators in the
realm of 1-dimensional (1D) conductors\cite{Degiorgi96,Gruner}, the
experimental situation at higher dimensions is less clear. Impossibility of
the Slater state in the 2D regime has been recently argued based on dynamical
cluster approximation calculations\cite{Moukouri01}. For 3D solids, spin
ordering alone usually generates an energy gap corrupting only a fraction of
the Fermi surface so that metallic conductivity persists\cite{Fawcett88}. In
this context the metal-insulator transition in the 3D pyrochlore Cd$_{2}%
$Os$_{2}$O$_{7}$ is exceptionally intriguing since transport and magnetic
properties across the MIT boundary appear to be in accord with the Slater
mechanism\cite{Mandrus01}. In this paper we report on the first spectroscopic
studies of the MIT in Cd$_{2}$Os$_{2}$O$_{7}$. Our analysis of the infrared
data reveals that the transition into the insulating state is driven solely by
the electronic interactions without significant involvement of the crystal
lattice. We discuss new facets of the spin-driven insulating state in a 3D material.

The pyrochlore Cd$_{2}$Os$_{2}$O$_{7}$ was first characterized by Sleight et
al.\cite{Sleight74}. The metal-insulator transition in the resistivity has
been found to occur at the same temperature $\simeq$ 226 K as the
antiferromagnetic transition in susceptibility measurements. No evidence of
concurrent structural changes were detected through X-ray diffraction (XRD)
analysis. Thorough examination of transport and magnetism in Cd$_{2}$Os$_{2}%
$O$_{7}$ has been recently reported by Mandrus et al.\cite{Mandrus01} with a
Slater picture delivering a coherent interpretation of all experimental data.
This particular mechanism in Cd$_{2}$Os$_{2}$O$_{7}$ may be favored by the
fact that Os$^{5+}$ is in the 5d$^{3}$ configuration so that $t_{2g}$ band is
near half filling.

The Slater transition is characterized by several hallmarks in the frequency
domain which so far remained unexplored in Cd$_{2}$Os$_{2}$O$_{7}$. Among
them, the specific temperature dependence of the energy gap as well as the
frequency dependence of the dissipative response at energies above the gap
edge that are distinct from the Mott-Hubbard case\cite{Thomas94}. Also the
analysis of the IR active phonon modes allows one to verify if a development
of charge ordering is concomitant with spin ordering. Despite the fact that IR
spectroscopy is ideally suited for directly studying the nature of the MIT in
solids, the extremely small size of Cd$_{2}$Os$_{2}$O$_{7}$ single crystals so
far has rendered these measurements impossible. Spectroscopic tools available
in our lab at UCSD are tailored for investigations of
microcrystals,\cite{Dordevic99} allowing us to fill voids in the experimental
picture of the insulating state in Cd$_{2}$Os$_{2}$O$_{7}$.%
\begin{figure}
[ptb]
\begin{center}
\includegraphics[
angle=270,width=3.6434in
]%
{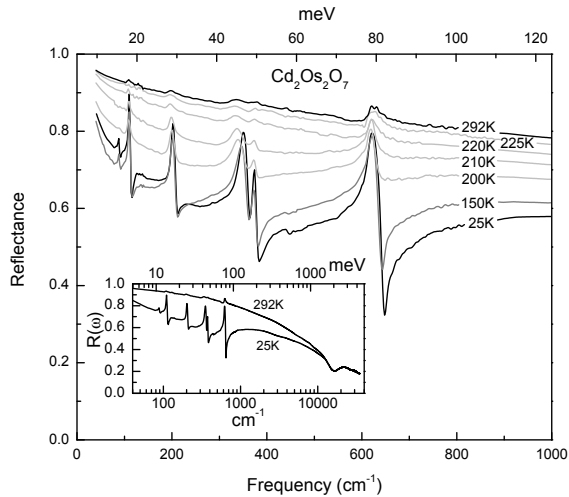}%
\caption{Reflectance of single crystal Cd$_{2}$Os$_{2}$O$_{7}$ at various
temperatures from 40 cm$^{-1}$ to 1000 cm$^{-1}$. Inset shows the entire
energy range characterized for room temperature and 25K on a log scale.}%
\label{fig1}%
\end{center}
\end{figure}

The near normal reflectance R$(\omega)$ of Cd$_{2}$Os$_{2}$O$_{7}$ was
measured from 40 cm$^{-1}$ to 14000 cm$^{-1}$ using a Fourier transform
spectrometer and from 12000 cm$^{-1}$ to 35000 cm$^{-1}$using a grating
monochromator. A test involving polarized light displayed no signs of
anisotropy, thus unpolarized radiation was used for a detailed study of the
temperature dependence. Samples were coated \textit{in situ} with gold or
aluminum and spectra measured from the coated surface were used as a
reference. This method, discussed previously in detail\cite{Homes93}, allows
one to reliably obtain the absolute value of the reflectance by minimizing the
errors associated with non-specular reflection and small sample size. We
inferred the complex conductivity $\sigma_{1}(\omega)+i\sigma_{2}(\omega)$ by
use of Kramers-Kronig (KK) analysis after extrapolating data to $\omega
\rightarrow0$ and $\omega\rightarrow\infty$. Various low frequency
extrapolations (Drude, Hagen-Rubens) were used; however the data did not
significantly depend on the particular method of low-$\omega$ extrapolation.
We employed the usual $\omega^{-4}$ dependence for the high-energy extension
of the data\cite{Wooten}.

Fig.\ref{fig1} shows the infrared reflectance taken between 25 K and room
temperature from an unpolished facet of Cd$_{2}$Os$_{2}$O$_{7}$ crystal with
dimensions less than 0.5x0.7 mm$^{2}$. The reflectance at room temperature is
high and is metallic in nature. R$(\omega)$ decreases with decreasing
temperature in the infrared region. Although the reflectance decreases
monotonically, it changes little near room temperature and at low
temperatures. Depression of R$(\omega)$ is most significant in the vicinity of
the MIT: between 225K and 150K. The absolute value of the reflectance for the
80K data is, within experimental accuracy, equivalent to that at 25K. As
temperature is lowered the free electron screening is gradually reduced
unveiling several strong phonons in the infrared region. Notably, all phonons
are still visible in the room temperature data. In the inset to Fig.\ref{fig1}
we plot the reflectance over the entire range characterized for room
temperature and 25K.%
\begin{figure}
[ptb]
\begin{center}
\includegraphics[
angle=270,width=3.6434in
]%
{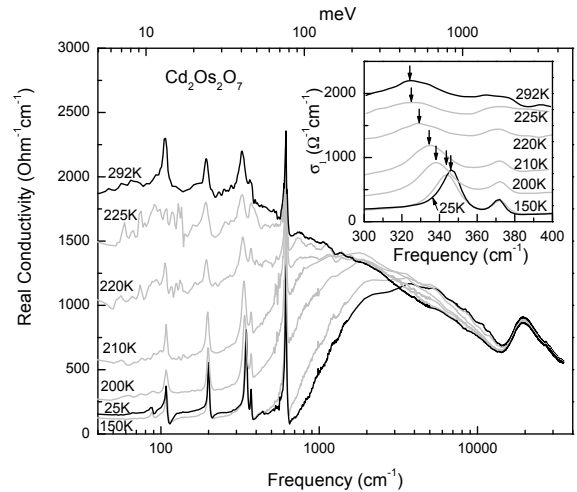}%
\caption{Real part of the optical conductivity as obtained by Kramers-Kronig
analysis on a log scale. A gap can be seen to develop continuously as the
temperature is reduced. The reduction in spectral weight occurring in the
infrared region is compensated by a shift to higher energies. The inset shows
the hardening Real part of the optical conductivity as obtained by
Kramers-Kronig analysis on a log scale. A gap can be seen to develop
continuously as the temperature is reduced. The reduction in spectral weight
occurring in the infrared region is compensated by a shift to higher energies.
The inset shows the hardening of the 347 cm$^{-1}$ phonon at low T.}%
\label{fig2}%
\end{center}
\end{figure}

\begin{figure}
[ptb]
\begin{center}
\includegraphics[
angle=270,width=3.6434in
]%
{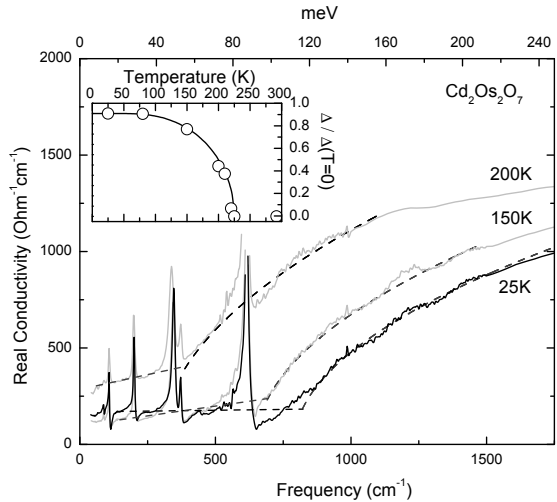}%
\caption{Infrared region of the real conductivity at $T<T_{MIT}$. The expected
theoretical frequency dependence of the gap edge ($\sigma_{1}(\omega
)\thicksim\omega^{1/2}$) is shown for each temperature as a dashed line. A
linear fit for the region below the gap edge is also shown. The intersection
of the two fits can be taken as a measure of the energy gap 2$\Delta$. The
inset shows the optical energy gap as determined by the method described above
(open circles), and the expected theoretical dependence (solid curve).}%
\label{fig3}%
\end{center}
\end{figure}

Fig.\ref{fig2} shows the real part of the conductivity $\sigma_{1}(\omega)$ at
different temperatures. The room temperature spectrum can be adequately
described with a simple Drude model $\sigma_{1}(\omega)=\sigma_{0}%
/(1+\omega^{2}\tau^{2})$ at least for $\omega<$ 1000 cm$^{-1}$ where
$\sigma_{0}$ is the DC conductivity and $\tau$ is the relaxation time. This
behavior is followed by a somewhat slower decrease than that prescribed by the
Drude form, and an interband feature at 22000 cm$^{-1}$. Several strong
phonons are visible in the IR region. As temperature is lowered one witnesses
the continuous opening of a gap with a drastic reduction of $\sigma_{1}%
(\omega)$ in the infrared. In the 25 K spectrum the intragap conductivity is
frequency independent (apart from the sharp phonon peaks). At $\omega$ above
the gap edge the spectrum reveals the $\omega^{1/2}$ dependence expected for a
Slater transition\cite{Thomas94,Millis01}. An intersection between the two
segments at 818 cm$^{-1}$ can be chosen as a quantitative measure of the
energy gap 2$\Delta$. The $\omega^{1/2}$ behavior can be recognized at higher
temperatures as well. In the latter spectra (Fig.\ref{fig3}) the intragap
region is best described with a linear dependence. The temperature dependence
of the intersection between the $\sigma_{1}(\omega)\propto\omega$ and
$\sigma_{1}(\omega)\propto\omega^{1/2}$ regions is plotted in the inset of
Fig.\ref{fig3}.%

\begin{figure}
[ptb]
\begin{center}
\includegraphics[
height=4.7124in,
width=3.6434in
]%
{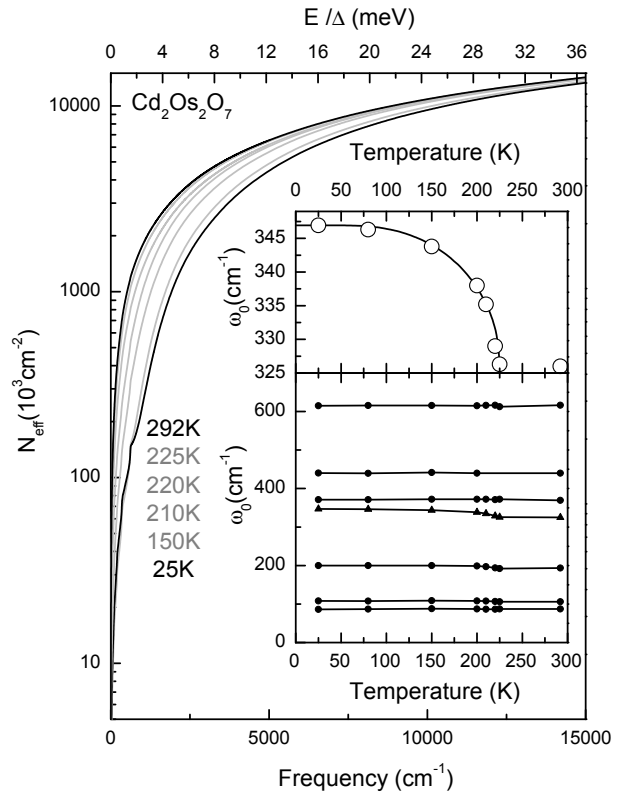}%
\caption{Effective spectral weight vs. cutoff frequency for temperatures as
indicated. The top axis is in energy units normalized by twice the gap energy.
The upper inset shows the temperature dependence of the 347 cm$^{-1}$phonon
and the BCS gap function is also plotted, (solid curve). The lower inset shows
the temperature dependence of all seven phonons.}%
\label{fig4}%
\end{center}
\end{figure}

It is instructive to characterize the development of the energy gap in
Cd$_{2}$Os$_{2}$O$_{7}$ through the spectra of the effective spectral weight
$N_{eff}(\omega)=\frac{120}{\pi}\int_{0}^{\omega}\sigma_{1}(\omega^{\prime
})d\omega^{\prime}$. The magnitude of $N_{eff}(\omega)$ depicted in
Fig.\ref{fig4}, is proportional to the number of carriers participating in the
optical absorption up to a cutoff frequency $\omega$, and has the dimension of
frequency squared. The significant reduction in spectral weight occurring in
the intragap region is transferred to the energy region above 3$\Delta$.
Interestingly, the spectral weight does not become completely exhausted until
16000 cm$^{-1}$ implying that the energy range as broad as $40\Delta
\simeq104k_{B}T_{MIT}$ is involved in the metal insulator transition.

Important insights into the origin of the insulating state in Cd$_{2}$Os$_{2}%
$O$_{7}$ may be reached through the analysis of the phonon spectra. The
pyrochlore structure belongs to the space group Fd\={3}m and reveals seven IR
active modes\cite{McCauley72,DeAngelis72}. We observe phonon peaks at 86
cm$^{-1}$, 108 cm$^{-1}$, 200 cm$^{-1}$, 347 cm$^{-1}$, 371 cm$^{-1}$, 440
cm$^{-1}$, and 615 cm$^{-1}$ in the 25K spectrum\cite{phonon-foot}. Both the
frequency position and the oscillator strength of all phonons (with the
exception of the 347 cm$^{-1}$ resonance) are independent of temperature
(lower inset of Fig. 4). The 347 cm$^{-1}$ mode assigned to the $O_{II}%
-Os-O_{II}$ bend\cite{McCauley72} shows weak hardening at $T<T_{MIT}$. The key
outcome of the examination of the phonon spectra in Figs.\ref{fig1}%
,\ref{fig2}, is that no new phonon modes appear at $T=T_{MIT}$ and the low-T
data for Cd$_{2}$Os$_{2}$O$_{7}$ displays only the 7 modes expected for the
ideal pyrochlore structure.%

The experimental evidence presented above suggests that the development of the
insulating state in Cd$_{2}$Os$_{2}$O$_{7}$ occurs without visible signs of
charge ordering. Examination of the IR-active phonons has proven to be one of
the most sensitive tests for the charge-ordered state. Quite commonly
additional lattice modes appear in the phonon spectra or a dramatic
redistribution of the spectral weight between several resonances takes place
provided the system reduces its symmetry in the charge-ordered
regime\cite{Buixaderas01}. We failed to detect any of these effects.
Furthermore, a detailed analysis of the x-ray diffraction data did not produce
any indications for structural changes at $T<T_{MIT}$\cite{Mandrus01}. These
observations allow us to conclude that the metal-insulator transition in
Cd$_{2}$Os$_{2}$O$_{7}$ is driven solely by electronic processes without
noticeable indications for involvement of the lattice. Another important fact
pertaining to the nature of the insulating state is the continuous development
of the energy gap in the electronic conductivity displayed in Fig.\ref{fig2}
which is consistent with the Slater picture of the MIT\cite{Matsubara53,Des
Cloizeaux59}. The second order transition is in accord with earlier specific
heat and magnetic susceptibility data\cite{Mandrus01}.

Further support for the Slater hypothesis in the context of the Cd$_{2}%
$Os$_{2}$O$_{7}$ data is provided by the electronic conductivity. We first
note that our data reveals several hallmarks of the Bardeen-Cooper-Schrieffer
(BCS) electrodynamics expected for systems with spin density
waves\cite{Gruner}, including the ratio of $2\Delta/k_{B}T_{MIT}\simeq5.1$
(expected in a modified BCS theory that takes into account the scattering of
electrons by phonons, as for the canonical SDW
Chromium)\cite{Barker68,Millis00} as well as a characteristic decline of the
gap value at non-zero temperatures \cite{Matsubara53,Des Cloizeaux59}. The
general form of the conductivity spectra is also consistent with the BCS
picture where type 2 coherence factors lead to an overshoot between the data
at $T\ll T_{MIT}$ and $T>T_{MIT}$ at frequencies above the gap. As pointed out
above, the behavior of the low-T spectra above the gap edge are adequately
described with the $\sigma_{1}(\omega)\thicksim\omega^{1/2}$ dependence, as is
expected for a Slater transition. This finding is important because the
$\omega^{1/2}$ dependence is distinct from the $\omega^{3/2}$ dependence
observed in the Hubbard limit\cite{Thomas94}.

Given the experimental evidence discussed above, the Slater mechanism emerges
as a viable model of the MIT in Cd$_{2}$Os$_{2}$O$_{7}$. Therefore, this
compound may be the first well documented case of a three-dimensional SDW
material\cite{Tokura95}, subject to further direct verification of the spin
structure. An unexpected feature of the antiferromagnetically-driven MIT is a
mismatch between the $T_{MIT}\simeq200K$ and the frequency range involved in
the redistribution of the spectral weight in the insulating state
$\Omega\simeq20000K$. Similar mismatch is commonly found throughout the
spectroscopic studies of the so-called pseudogap state in high-$T_{c}$
superconductors\cite{Basov02}, in which antiferromagnetic fluctuations are
perceived as a likely cause of the pseudogap state. Finally, it is worth
mentioning that other pyrochlore compounds reveal very different properties at
the verge of the metal-insulator transition. For instance, the transition to
the insulating regime in the closely related Tl$_{2}$Ru$_{2}$O$_{7}$ system
appears to be of first order and additionally is accompanied by charge
ordering effects judging from the transformations of the phonon
spectra\cite{Lee01}. It is yet to be determined what microscopic factors
define the peculiar character of the MIT in Cd$_{2}$Os$_{2}$O$_{7}$.

Work performed at UCSD was sponsored by the U.S. Department of Energy under
Contract No. DE-FG03-00ER45799. Work at Oak Ridge National Laboratory is
managed by UT-Battelle, LLC, for the U.S. Department of Energy under Contract
No. DE-AC05-00OR22725.

\bigskip

\begin{thebibliography}{99}
\bibitem {phonon-foot}Of these seven phonons, six can easily be resolved in
$\sigma_{1}(\omega)$ shown in Fig.\ref{fig2} , and a small hard to see phonon
located at 440 cm$^{-1}$ can be resolved easiest in Fig.\ref{fig1} in the
lower temperature data sets.
\end{thebibliography}
\end{document}